\documentclass[final,5p,times,twocolumn]{elsarticle} 
\usepackage{lineno,hyperref}

\journal{Nuclear Physics B}
\begin{document}

\begin{frontmatter}

\title{An automated QC Station for the Calibration of the Mu2e Calorimeter Readout Units}

\author[1]{C. Bloise}
\author[1]{S. Ceravolo}
\author[2]{F. Cervelli}
\author[1]{F. Colao }
\author[1]{M. Cordelli}
\author[1]{G. Corradi}
\author[2]{S. Di Falco}
\author[1]{E. Diociaiuti}
\author[2]{S. Donati}
\author[2]{C. Ferrari}
\author[1]{R. Gargiulo}
\author[4]{A. Gioiosa}
\author[1]{S. Giovannella}
\author[2]{V. Giusti}
\author[1]{D. Hampai}
\author[1]{F. Happacher}
\author[1]{M. Martini}
\author[1]{S. Miscetti}
\author[2]{L. Morescalchi}
\author[1]{D. Paesani}
\author[2]{D. Pasciuto}
\author[2]{E. Pedreschi}
\author[2]{F. Raffaelli}
\author[1]{E. Sanzani\corref{cor1}}
\ead{elisa.sanzani@lnf.infn.it}
\cortext[cor1]{Corresponding author}
\author[1]{I. Sarra}
\author[3]{A. Saputi}
\author[2]{F. Spinella}
\author[2]{A. Taffara}

\affiliation[1]{Laboratori Nazionali di Frascati dell'INFN, Frascati, Italy}
\affiliation[2]{INFN - Sezione di Pisa, Italy} 
\affiliation[3]{INFN - Sezione di Ferrara, Italy}
\affiliation[4]{Università degli Studi di Tor Vergata, Italy}

\vspace{-1 cm}
\begin{abstract}
The Mu2e calorimeter will employ Readout Units, each made of two Silicon Photomultipliers arrays and two Front End Electronics boards. To calibrate them, we have designed, assembled and put in operation an automated Quality Control (QC) station. Gain, collected charge and photon detection efficiency are evaluated for each unit. 
In this paper, the QC Station is presented, in its hardware and software aspects, summarizing also the tests performed on the ROUs and the first measurement results.
\end{abstract}

\end{frontmatter}

\nolinenumbers

\section{The Mu2e Electromagnetic Calorimeter Readout}

\begin{figure}[htb]
    \centering
    \includegraphics[scale=0.35]{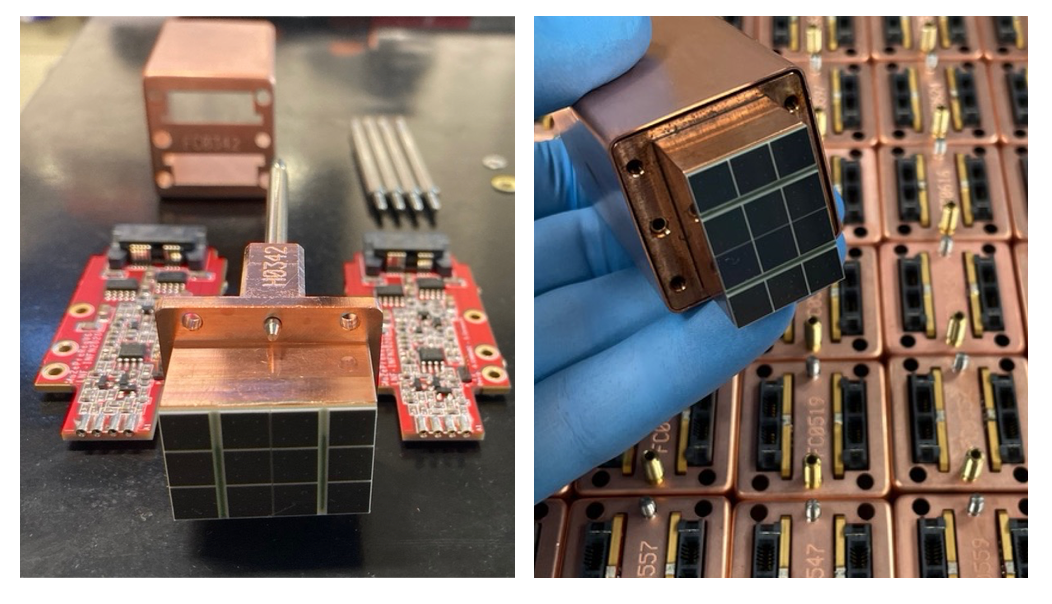}
    \caption{Picture of a ROU. On the left a disassembled ROU is show, allowing to see the FEE boards. On the right some assembled ROUs are shown.}
    \label{fig:foto_rous}
\end{figure}

The Mu2e experiment will search for the charged lepton flavour violation via the conversion process: $\mu^- \;^{27}{\rm Al} \to e^- \; ^{27}{\rm Al}$, with the aim of improving the current sensitivity on the ratio between the conversion and capture event rates by four orders of magnitude, reaching a sensitivity of $8\times 10^{-17}$ at 90\% CL \cite{bernstein2019mu2e}.
The Mu2e calorimeter is formed by two annular disks, each containing 674 undoped CsI crystals.
It will have to work in a $10^{-4}\,$Torr vacuum, a $1\,$T magnetic field and in a very harsh radiation environment \cite{atanov2018mu2e}. It has to satisfy strict requirements on time, $\sigma_{t}<500\,$ps,  space, $\sigma_{x}<1\,$cm and energy resolution $\sigma_{E}/E<10\%$, for 100 MeV electrons. The calorimeter is scheduled to be completed by the end of 2023.
Each CsI crystal is readout by two UV-extended Hamamatsu Silicon Photomultipliers, each made by a 2$\times$3 matrix of 6$\times6\,{\rm mm^{2}}$ monolithic cells with a total gain of $\mathcal{O}(10^6)$. Two Mu2e SiPMs glued on a copper holder and two independent Front End Electronics (FEE) boards form a Readout Unit (ROU) (Fig. \ref{fig:foto_rous}). 

\section{The LNF Quality Control Station for the Readout Units}
\begin{figure}[htb]
    \centering
    \includegraphics[scale=0.35]{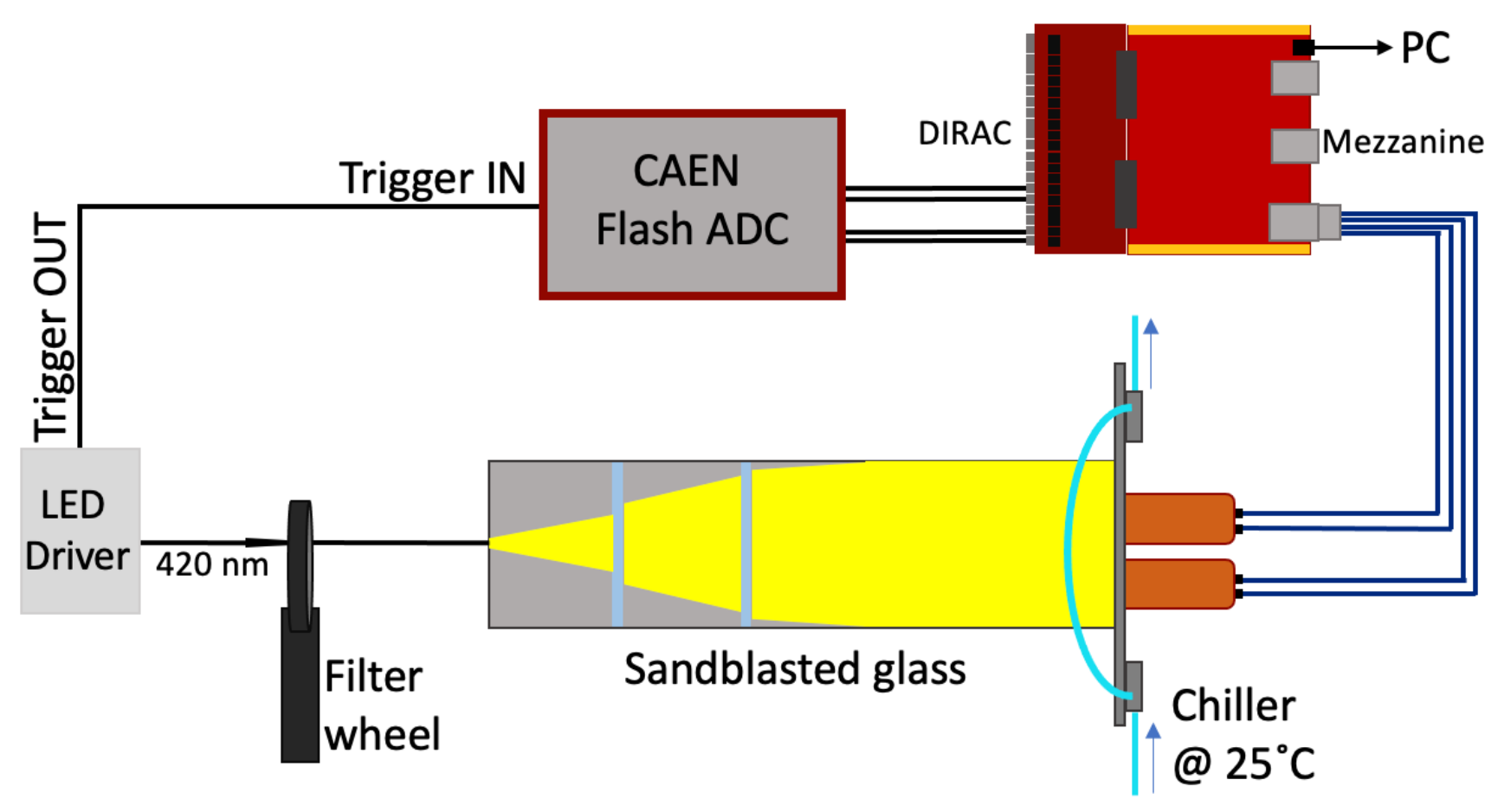}
    \caption{Schematics of the LNF ROU QC Station.}
    \label{fig:schema_QC}
\end{figure}

To qualify and calibrate the $\mathcal{O}(1500)$ assembled ROUs, we have designed, realized and put in operation an automated QC station at LNF (Laboratori Nazionali di Frascati of INFN). 
In this station, the crystal luminescence light is mimicked by a 420 nm LED which illuminates the sensors through an attenuating set of filters held by an automatic wheel. This allows to obtain nine different light intensities. A metal box ensures light-tightness and houses two sandblasted glass layers to uniformly diffuse the light on the SiPMs. The Al plate holding the ROUs is temperature stabilized at 25 °C via a chiller. In Fig. \ref{fig:schema_QC} a scheme of the station is shown. 
A Mu2e Mezzanine board, USB connected to a PC, controls the supplied high voltage to the SiPMs and provides the analog signals for digitization. The data acquisition, triggered by the LED driver signal, is based on a CAEN waveform digitizer. Two power supplies are used at the station, one Low Voltage to power the Mezzanine and one High Voltage to provide the SiPM bias voltage, both remotely controlled.

Data acquisition is performed on two ROUs at the same time. Since the SiPMs will sustain radiation damage during detector operation, we plan to decrease their bias voltage ($V_{bias}$) up to $4\,$V to limit the leakage current. To study the SiPM response dependence on $V_{bias}$, a scan is performed around the SiPM operational voltage, $V_{op} \sim 165\,$V, from $V_{op}-4\,$V to $V_{op}+2\,$V, in 1V steps. For every voltage value,  $10^4$ events per wheel position are acquired. Underlying C++ and Python programs control the data acquisition and thanks to the high level of parallelization of the analysis, a full scan is performed in 7 minutes. 
This setup allows to determine the SiPM gain, photon detection efficiency (PDE) and the collected charge for each voltage step to study their $V_{bias}$ dependence. A sensor on the FEE boards allows to monitor the detector temperature and thus evaluates the temperature dependence of the gain. 
\section{Calibration and Quality Control results}
\begin{figure}[htb]
    \centering
    \includegraphics[scale=0.22]{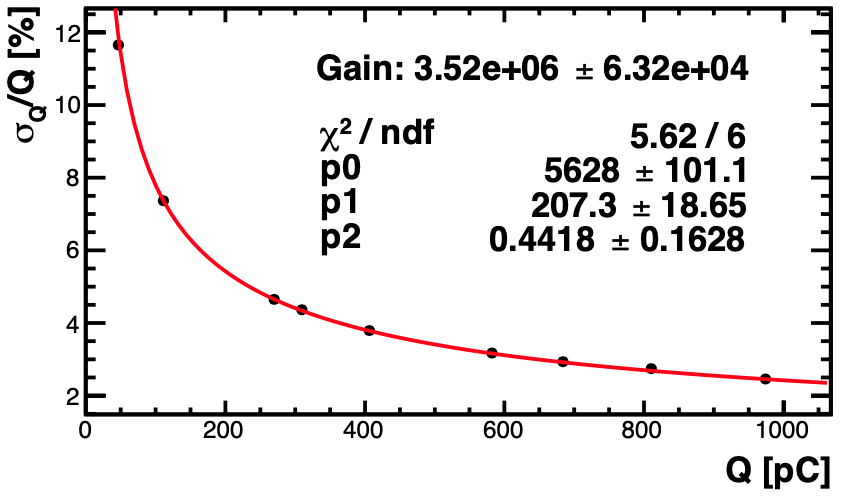}
    \caption{$\sigma_Q/Q$ results for a scan in light intensity for the 9 filters at $V_{op}$. Notice the percentage scale in the plot.}
    \label{fig:plot_scan}
\end{figure}

\begin{figure}[htb]
    \centering
    \includegraphics[width=\linewidth]{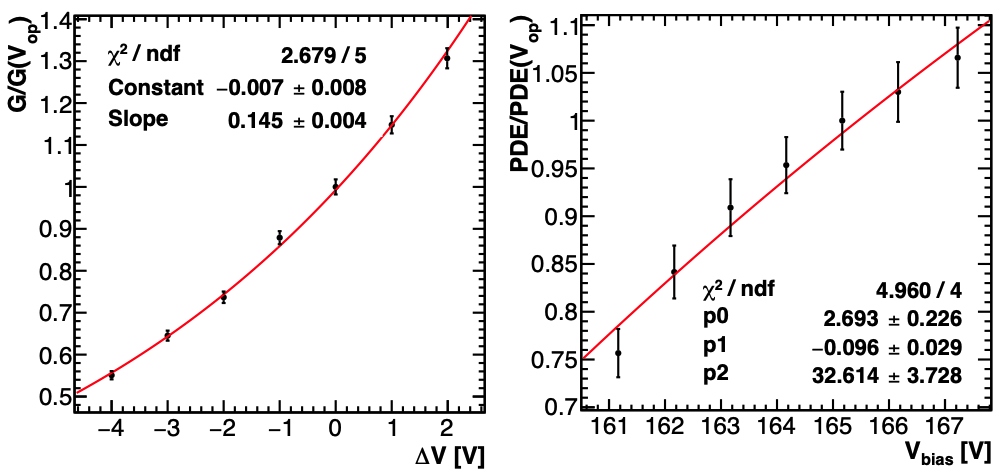}
    \caption{Behaviour of gain (left) as a function of $\Delta V$ and PDE (right) as a function of $V_{bias}$. Gain is fitted with an exponential and the PDE with $\mathrm{PDE(V)=p_0\cdot[1-(p_1\cdot V\cdot exp(-p_2/\sqrt{V})^{-2}]}$ \cite{Gallina_2019}.
    }
    \label{fig:plot_scan_fit}
\end{figure}

An example of a scan result over the nine filters at one voltage value is  shown in Fig. \ref{fig:plot_scan}. From the measurement of $\sigma_Q/Q$ it is possible to extract the SiPM gain through the relation:
$  \frac{\sigma_Q}{Q}(Q)= \sqrt{\frac{p_0}{Q}+\frac{p_1^2}{Q^2}+p_2^2}$.
The gain is evaluated as $G=p_0/q_e$ with $q_e=1.6\cdot10^{-7}\,$pC. 
This result is obtained for each supplied voltage value so that the gain as a function $V_{bias}$ can be determined. The obtained behaviour is shown in Fig. \ref{fig:plot_scan_fit} as a function of the overvoltage $\Delta V = V_{bias} - V_{op}$.

\begin{figure}[htb]
    \centering
    \includegraphics[width=\linewidth]{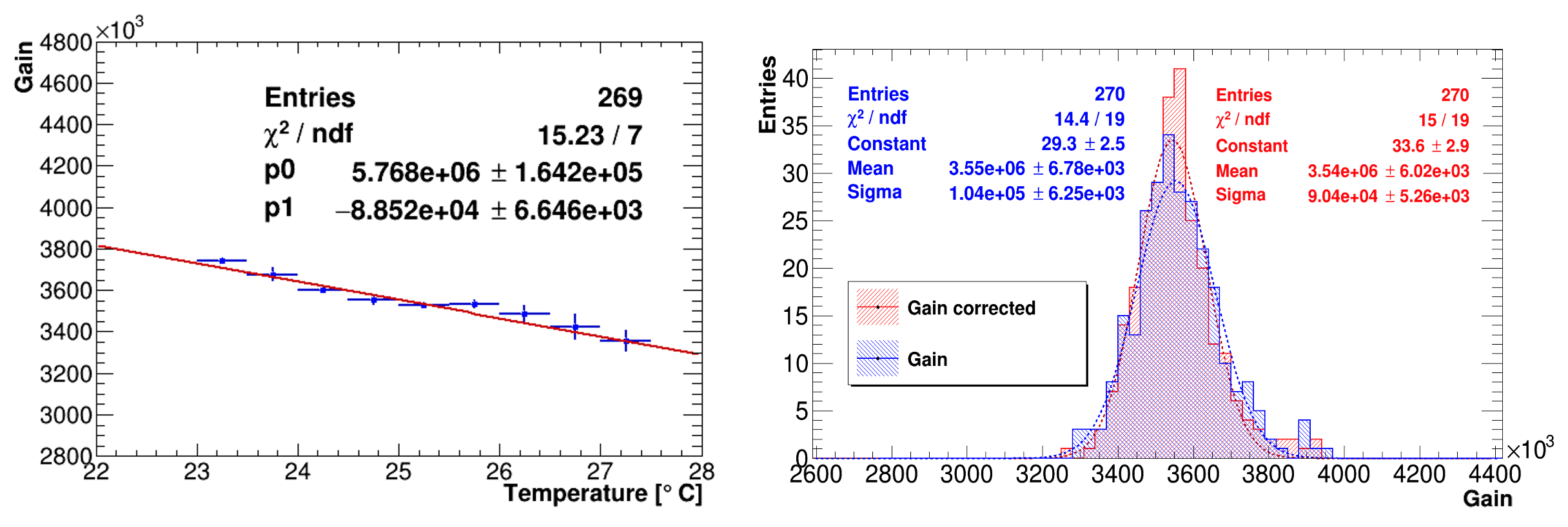}
    \caption{Left: Temperature profile of the gain. Right: Gain distribution with and without the temperature correction applied.}
    \label{fig:plot_tempprof}
\end{figure}

The statistical reproducibility of the gain measurement  has been tested and it is evaluated to be better than 2\%. 
The temperature dependence has also been evaluated and it has been found that the breakdown voltage variation due to a temperature increase causes a gain decrease of $1.6\%/^{\circ}{\rm C}$. This value has been extracted from the temperature profile of the gain in a range of $\pm2\,^{\circ}{\rm C}$ around $25\,^{\circ}{\rm C}$, as shown in Fig. \ref{fig:plot_tempprof} (left). 
To build a uniform ROU database and to have comparable data, we have corrected the gain values to $25\,^{\circ}{\rm C}$. The gain distribution before and after temperature correction is shown in Fig. \ref{fig:plot_tempprof} (right). The latter appears narrower and with a smaller population at the tails.
 
\section{Conclusions}
The Mu2e Calorimeter has strict requirements in terms of performance and stability. To ensure these requirements are met, a full characterization of the ROU parameters has been implemented. The QC station allows to perform a HV scan of 2 ROUs in 7 minutes. The dependence of gain, charge and PDE on the SiPM overvoltage and on temperature has been studied. The average gain value at operational voltage is $3.6\times10^6$, with a spread along production of the order of 3\%. The ROUs scan is ongoing at the moment of writing and it is expected to be completed by the end of 2022. 

\section{Acknowledgements}
We are grateful for the vital contributions of the Fermilab staff and the technical staff of the participating institutions. This work was supported by the US Department of Energy; Istituto Nazionale di Fisica Nucleare, Italy; the Science and Technology Facilities Council, UK; the Ministry of Education and Science, Russian Federation; the National Science Foundation, USA; the Thousand Talents Plan, China; the Helmholtz Association, Germany; and the EU Horizon 2020 Research and Innovation Program under the Marie Sklodowska-Curie Grant Agreement Nos. 101003460, 101006726, 734303, 822185, and 858199. This document was prepared by members of the Mu2e Collaboration using the resources of Fermilab, a U.S. Department of Energy, Office of Science, HEP User Facility. Fermilab is managed by Fermi Research Alliance, LLC (FRA), acting under Contract No. DE-AC02-07CH11359.


\bibliography{mybibfile}

\end{document}